\documentclass[letter]{aa}
\usepackage{graphicx,epsfig,fancyhdr,rotating,amsmath,natbib}
\usepackage{txfonts,url}
\usepackage{color}
\def\etal{{et al.}}
\def\ie{{i.e. }}

\def\arcsec{$^{\prime\prime}$}

\begin{document}
   \title{Signatures of Alfv\'{e}n waves in the polar coronal holes as seen by EIS/Hinode}

   \author{D. Banerjee\inst{1} \and D. P\'erez-Su\'arez\inst{2} \and J.G. Doyle\inst{2} }

\offprints{dipu@iiap.res.in}

\institute{Indian
Institute of Astrophysics, Bangalore 560034, India \and Armagh Observatory, College Hill, 
Armagh BT61 9DG, N.Ireland}
\date{}
\abstract 
{We diagnose the properties of the plume and interplume regions in a polar 
coronal hole and the role of waves in the acceleration of the solar wind.}
{We attempt to detect whether Alfv\'{e}n waves are present in the polar coronal holes 
through variations in EUV line widths.}
{Using spectral observations performed over a polar coronal hole region with 
the EIS spectrometer on Hinode, we study the variation in the line width and 
electron density as a function of height. We use the density sensitive line pairs of Fe~{\sc xii} 186.88 \AA\ \& 195.119~\AA\ 
and Fe~{\sc xiii} 203.82 \AA~\& 202.04 \AA~.}
{For the polar region, the 
line width data show that the nonthermal line-of-sight velocity 
increases from $26 \ {\rm km \; s}^{-1}$ at 10\arcsec\ above the limb to $42
\;{\rm km \; s}^{-1}$ some 150\arcsec\ (\ie $\sim$110,000 km) above the limb. 
The electron density shows a decrease from $3.3\times10^9 \ {\rm cm}^{-3}$ to 
$1.9\times10^8 \ {\rm cm}^{-3}$ over the same distance. } 
{These results imply that the nonthermal velocity is inversely proportional 
to the quadratic root of the electron density, in excellent agreement with 
what is predicted for undamped radially propagating linear Alfv\'{e}n waves. Our data 
provide signatures of Alfv\'{e}n waves in the polar coronal 
hole regions, which could be important for the acceleration of the solar wind.}

\titlerunning{Alfv\'{e}n waves in the polar coronal holes}
\authorrunning{Banerjee \etal}

\maketitle
\section{Introduction}

Over the past decade, data from Ulysses show the importance of the polar coronal holes, 
particularly at times near solar minimum, for the acceleration of the 
fast solar wind. Acceleration of the quasi-steady, high-speed solar wind 
requires additional energy in the 
supersonic region of the flow. It has been shown theoretically that 
Alfv\'{e}n waves from the Sun can accelerate the solar wind to these high 
speeds. Until now this is the only mechanism shown to enhance 
the flow speed of the basically thermally driven wind to the high flow 
speeds observed in interplanetary space. The Alfv\'{e}n speed in the 
corona is quite high, so Alfv\'{e}n waves can carry a significant energy 
flux even for low wave energy density. Direct observations of Alfv\'{e}n waves has gained  
momentum after the launch of the Hinode satellite. Recent reports of  
detections of low-frequency ($<$ 5 mHz), propagating Alfv\'{e}nic motions 
in the solar corona \citep[][from coronagraphic observations]{2007Sci...317.1192T}  and chromosphere 
\citep{2007Sci...318.1574D} and their relationship with chromospheric 
spicules observed at the solar limb \citep{2007PASJ...59S.655D} with the 
Solar Optical Telescope \citep[SOT;][]{2008SoPh..249..167T} on Hinode 
\citep{2007SoPh..243....3K} have widened interest in the subject.
If MHD waves play an important role in accelerating high-speed streams and coronal heating, 
then this should be observed via a broadening of spectral lines, increasing with radial distance in 
the inner corona.

There have been several off-limb spectral line observations to 
search for Alfv\'{e}n wave signatures. Measurements of ultraviolet 
\ion{Mg}{x} line widths during a rocket flight showed an increase in 
width with height to a distance of 70,000~km \citep{1990ApJ...348L..77H}. Also, coronal hole \ion{Fe}{x} 
spectra taken at Sacramento Peak Observatory    
showed an increase in the line width with height \citep{1994SSRv...70..373H}. 
SUMER/SoHO \citep{1995SoPh..162..189W}  
were used to record the off-limb, height-resolved spectra of an Si~{\sc viii} 
density sensitive line pair, in  equatorial coronal regions 
\citep{1998SoPh..181...91D,2005A&A...435..733W} and in polar coronal holes 
\citep{1998A&A...339..208B,2004A&A...415.1133W}. The measured variation in line width with 
density and height supports undamped wave propagation in coronal structures. This 
was strong evidence of outwardly propagating 
undamped Alfv\'{e}n waves in the corona, which may contribute to coronal 
heating and high-speed solar wind in the case of coronal holes. We revisit the subject with the 
new EIS instrument on Hinode and compare with our previous results as 
recorded by SUMER/SoHO. In the present study, we also make a full scan in the 
off-limb polar region to study the differences between 
the plume and interplume regions and to address the question as to whether 
the plume or the interplume is the preferred channel for the acceleration 
of the wind. Analyses of SUMER  \citep[e.g.][]{2000SoPh..194...43B, 2003ApJ...588..566T} and 
UVCS  \citep[e.g.][]{2000ApJ...531L..79G}  data have shown that the width 
of UV lines in interplumes is greater than in plume regions, indicating 
that interplumes as the site where energy is preferentially deposited and, 
possibly, the fast wind emanates.
% Moreover, \cite{2000ApJ...531L..79G}, from
%a Doppler dimming analysis of the O~{\sc vi} 1032 and 1037 \AA\ lines at 
%1.7 R$_{\sun}$, found a larger outflow speed in interplume than in plumes. 
In this letter we concentrate on the density diagnostic capability of the density-sensitive line pairs of \ion{Fe}{xii}~186.88~\AA\ and 195.119~\AA\ \citep{2009A&A...495..587Y} and \ion{Fe}{xiii} 203.82~\AA\ and 202.04~\AA\ \citep{2009ApJ...692.1294W} for the off-limb coronal hole region. From a study of the variation of line width of the strongest line within our spectrum, we try to find signatures of propagating Alfv\'{e}n waves.
\begin{figure}
\centering
%\vspace*{-0.2cm}
\includegraphics[bb= 45 55 520 308, clip= true, width=8cm]{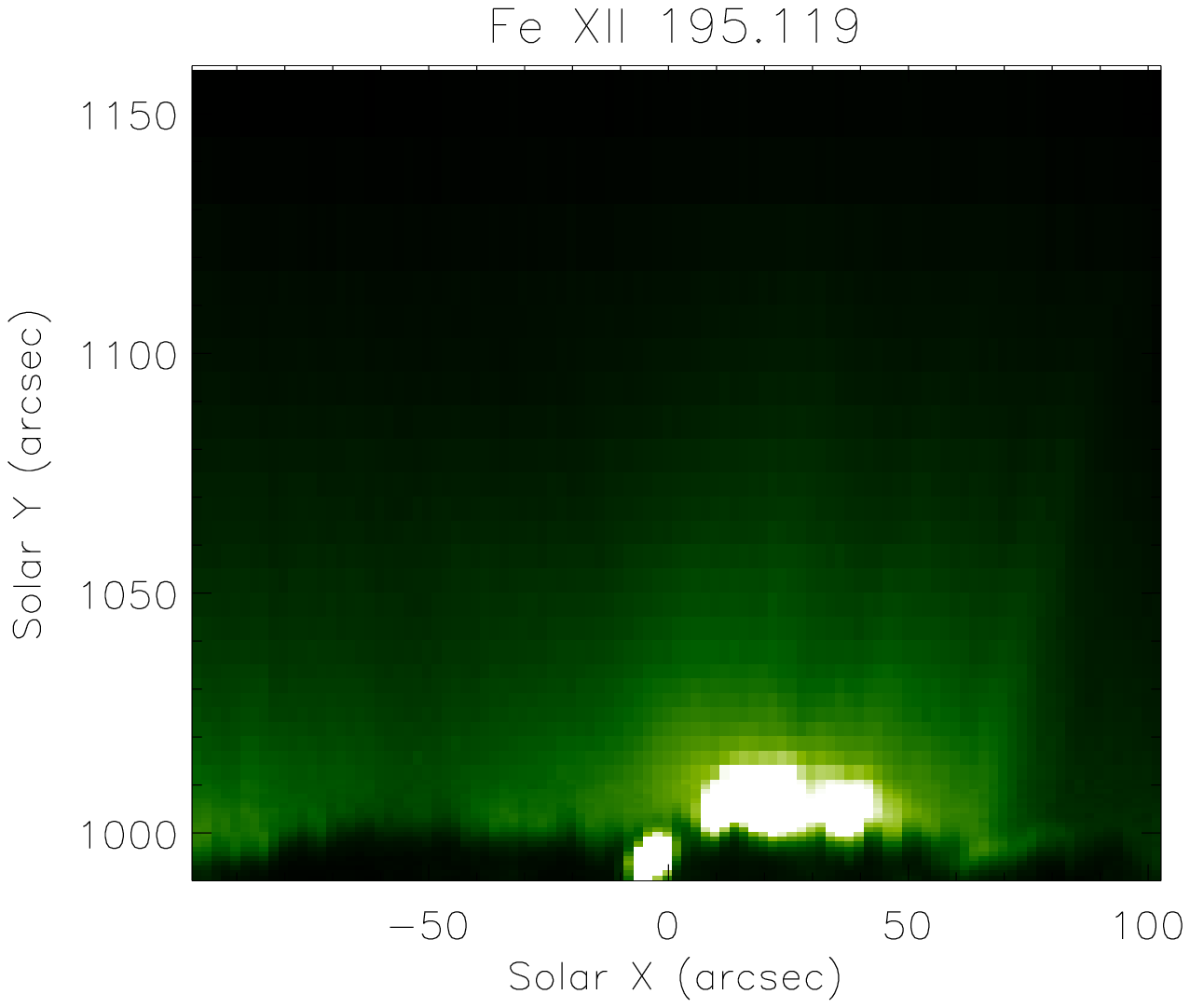}
\includegraphics[bb= 45 50 520 308, clip= true, width=8cm]{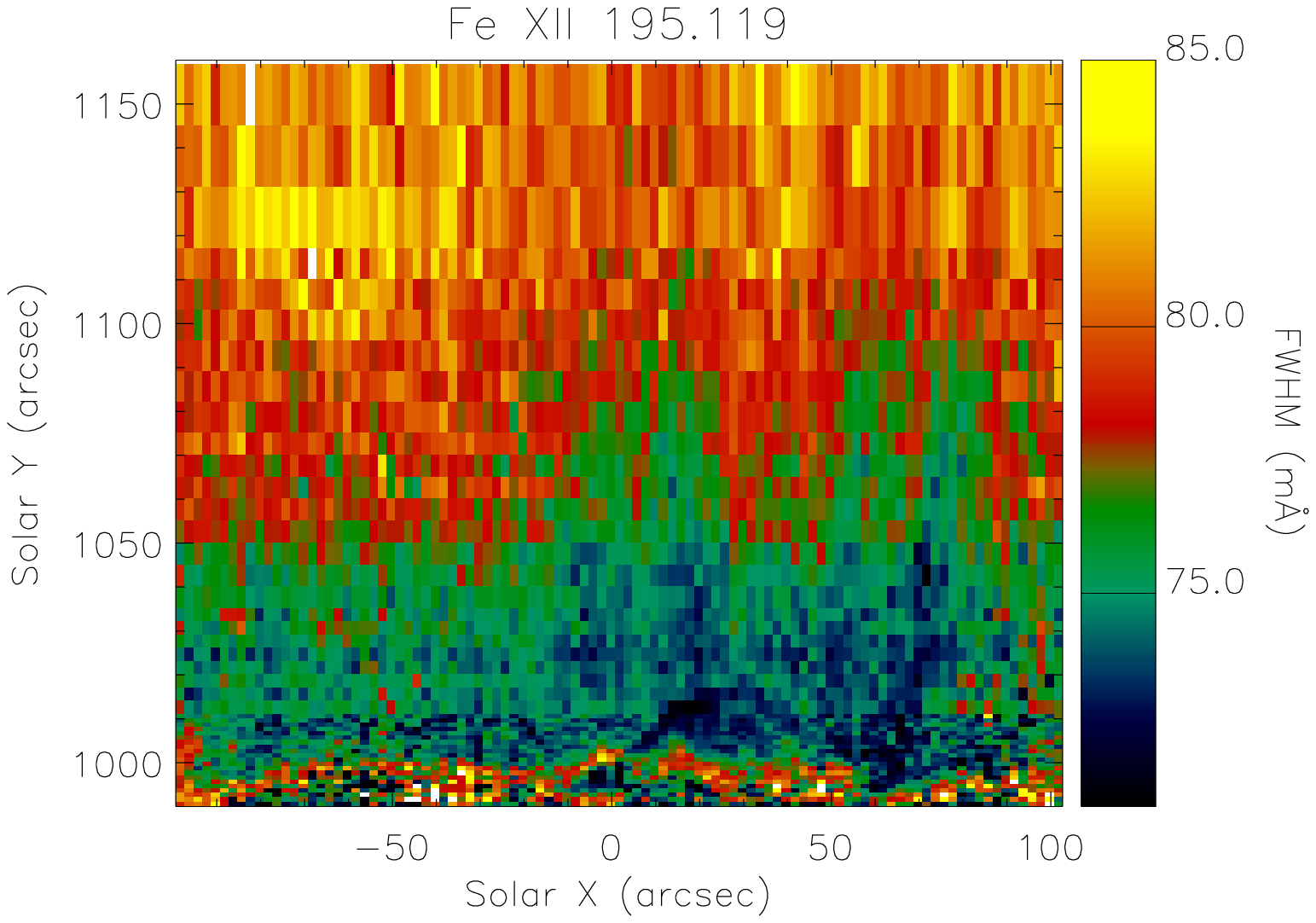}\vspace*{-0.1cm}
\includegraphics[bb= 45 0 520 308, clip= true, width=8.cm]{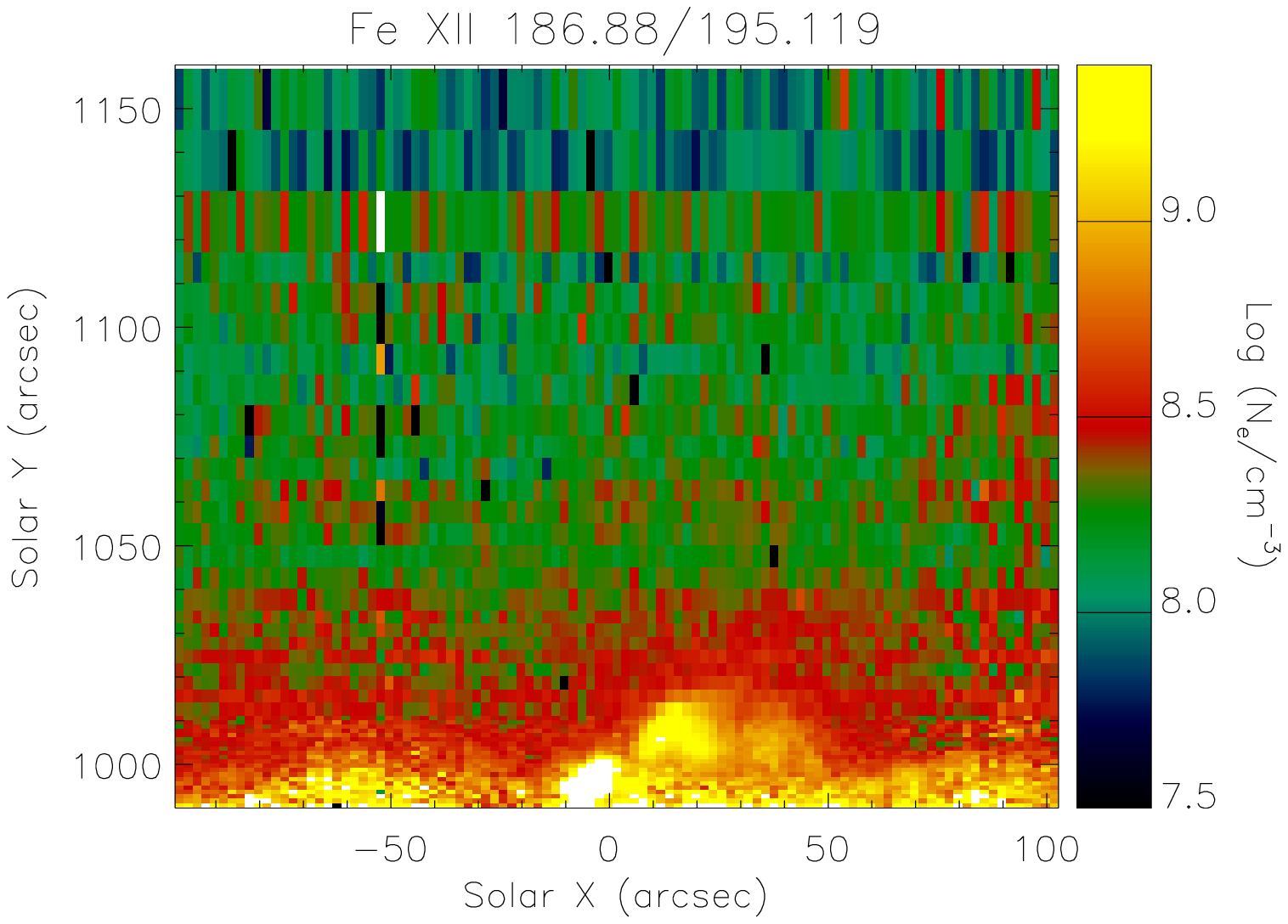}
\caption {Images of the north polar coronal hole off-limb region. From top to bottom:
intensity, width, and density obtained from  \ion{Fe}{xii} lines. } \label{fig:map}
\end{figure}
%%%%%%%%%%%%%%%%%%%%%%%%%%%%%%%%%%%%%%%%%%%%%%%%%%%%%%%%%%%%%%%%
\section{Observation and data reduction}
EIS onboard Hinode observes high-resolution spectra in two
wavelength bands 170-211 \AA\ and 246-292 \AA\ with a spectral
resolution of 0.0223 \AA\ per pixel \citep{2007SoPh..243...19C}.
We observed the north polar coronal hole on and off the limb with the 
2\arcsec\ slit on 10 October 2007 (see left panel of Fig.~\ref{fig:densitymapfull}, available online only, for a context image). The raster was run for more than 
four hours with 101 exposures of 155~s, 
covering an area of 201.7\arcsec$\times$512\arcsec. We used the study arm\_raster\_ar, which was 
designed with nine windows containing lines from {\ion{Fe}{viii}, \ion{Fe}{x}, \ion{Fe}{xi},
\ion{Fe}{xii}, \ion{Fe}{xiii}, \ion{Fe}{xiv}, \ion{Fe}{xv}, \ion{He}{ii},
\ion{Mg}{vi}, \ion{O}{v}, \ion{O}{vi}, \ion{Si}{vii}, and \ion{Al}{ix}. All data have been reduced and calibrated with the standard procedures as given 
in the SolarSoft(SSW)\footnote{\url{http://www.lmsal.com/solarsoft/}} library. 
These standard subroutines include dark current subtraction, cosmic ray 
removal, flat field, hot pixel correction, and absolute calibration. We use the density sensitive line pairs of \ion{Fe}{xii}
186.88~\AA\ and 195.119~\AA\ 
\citep{2009A&A...495..587Y}  and \ion{Fe}{xiii} 203.82~\AA\ and 202.04~\AA\ \citep{2009ApJ...692.1294W} for density diagnostics and use the strongest line of  \ion{Fe}{xii}~195.119~\AA\ for the line width study.
 These lines are formed in the temperature range of $T$ = 1 to 2 MK.
One can see from the 
CHIANTI\footnote{\url{http://www.solar.nrl.navy.mil/chianti.html}}
\citep{2006ApJS..162..261L} atomic data base that these line pairs
have a very good density sensitivity between 7.0 $\leq$ $\log$
N$_{e}$ (cm$^{-3}$) $\leq$ 10.0. 

The \ion{Fe}{xii} 195.12 \AA\ was fitted using a double Gaussian  
\citep[see][]{2009A&A...495..587Y}, considering the two main transitions at 
195.119 and 195.179~\AA~(with a 5\% contribution on average to the stronger component).  The two transitions for \ion{Fe}{xii} at
186.854 and 186.887 were fitted with a single Gaussian at 186.88~\AA. For \ion{Fe}{xiii}, the new CHIANTI updated version was used,  which incorporates new Fe ion models as described in \cite{2009ApJ...692.1294W}. To address the issue of weaker signal-to-noise 
as we go off-limb, a variable binning in the radial direction was 
performed, where after 1010\arcsec\ we initially binned   
over 27 pixels, finishing by binning over 42 pixels far off-limb. For
\ion{Fe}{xiii} we also binned over 5 spectra in the X direction.
The ``grating tilt'' reported by \citet{2009A&A...495..587Y}, which clearly affects line
ratios, was corrected by assuming that the tilt is linear and using the
value found by \citet{2009A&A...495..587Y} after our own comparison.

%%%%%%%%%%%%%%%%%%%%%%%% FIG Y cut %%%%%%%%%%%%%%%%%%%%%%%%%%%%%%%%%%%%%%%%%
\begin{figure}
\centering
\includegraphics[width=8cm]{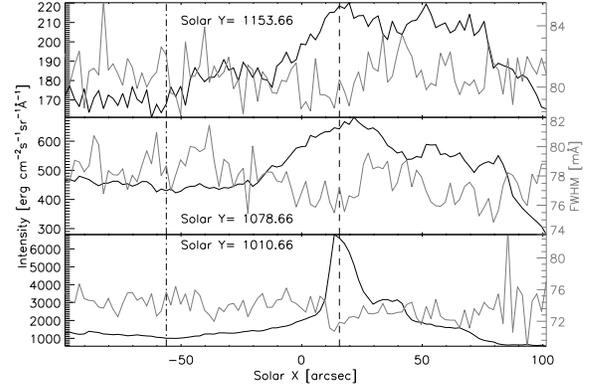}
\caption {A plot of the  intensity (solid lines) and FWHM (grey lines) as 
recorded by \ion{Fe}{xii}~195 \AA\ along three strips tangent to the limb at different 
heights to illustrate the anti-correlation between 
intensity and velocity. The vertical lines show the position of the plume 
(dashed) and interplume (dot-dashed). The limb corresponds to Y=1003, as decided from the intensity peak of a low temperature line, \ion{Si}{vii}.} \label{fig:Ycut}
\end{figure}
%----------------------------------------------------------------------------
\section{Results}
The spectral line profile of an optically thin coronal emission line results 
from the thermal broadening caused by the ion temperature $T_{i}$, as well as
 by small-scale unresolved nonthermal motions and/or 
unresolved flows. Assuming that the instrumental profile can be expressed as Gaussian, the expression for the full width half
maximum (FWHM) is given as
\begin{equation}
FWHM\,=\,\left[W^{2}_{inst} +
4ln2\,\left(\frac{\lambda}{c}\right)^2\,\left(\frac
{2\,k\,T_{i}}{M_{i}}\,+\,{\xi}^2\right)\right]^{1/2}
\end{equation}
where $T_{i}$, $M_{i}$, and $\xi$ are respectively the ion temperature, ion 
mass, and nonthermal velocity, while $W_{inst}$ is the instrumental width (for 
the 2\arcsec\ slit we use 62 m\AA\ \footnote{\url{http://msslxr.mssl.ucl.ac.uk:8080/eiswiki/Wiki.jsp?page=2EISSlit}}), with the ion temperature assumed equal 
to the peak in the line formation temperature as given by CHIANTI.
%%%%%%%%%%%%%%%%%%%%%%%% FIG density comparison for Fe 12 and Fe 13 with height %%%%%%%%%%%%%%
\begin{figure}
\centering
\includegraphics[bb= 0 45 520 340, clip= true, width=8cm]{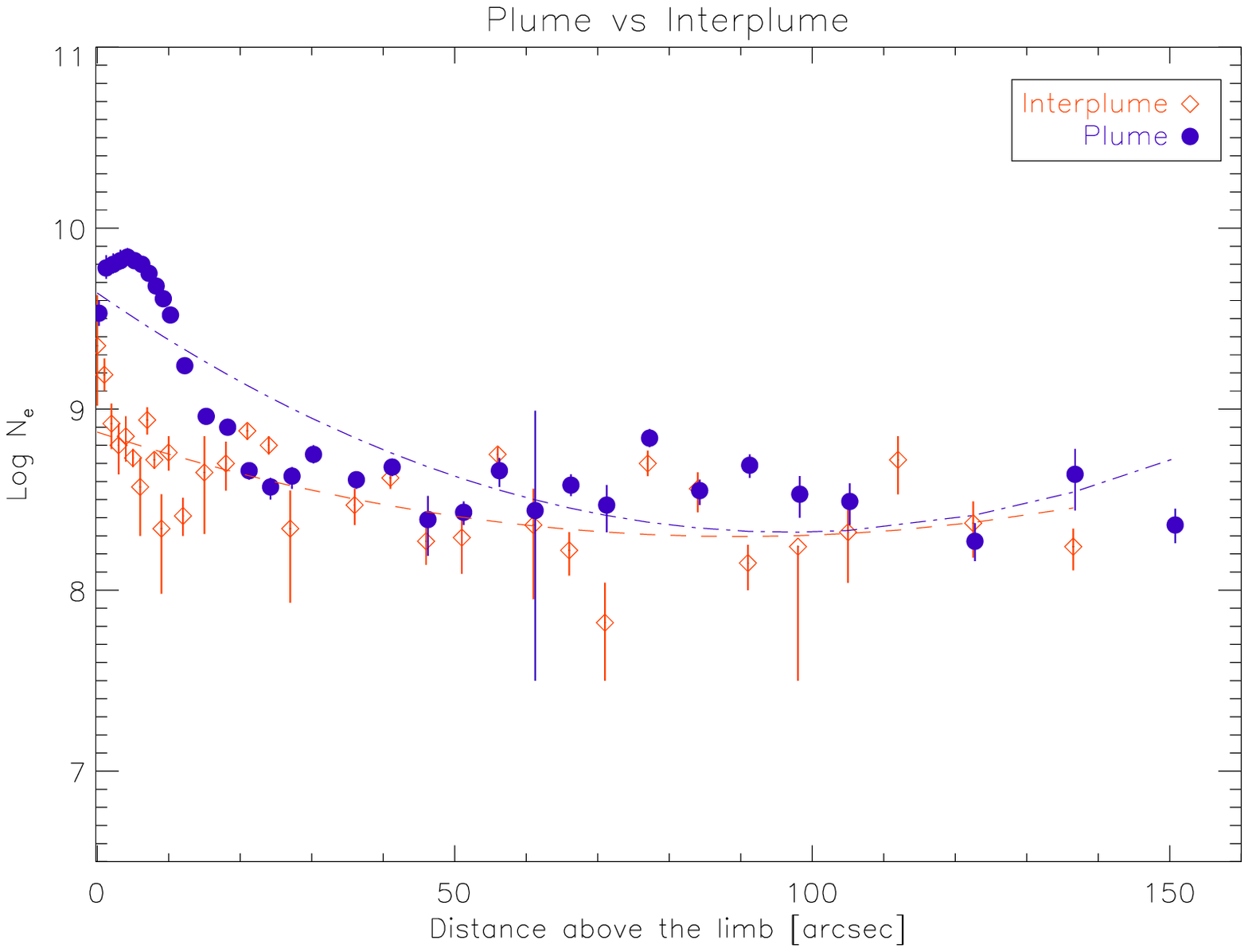}
\includegraphics[bb= 0 45 520 340, clip= true, width=8cm]{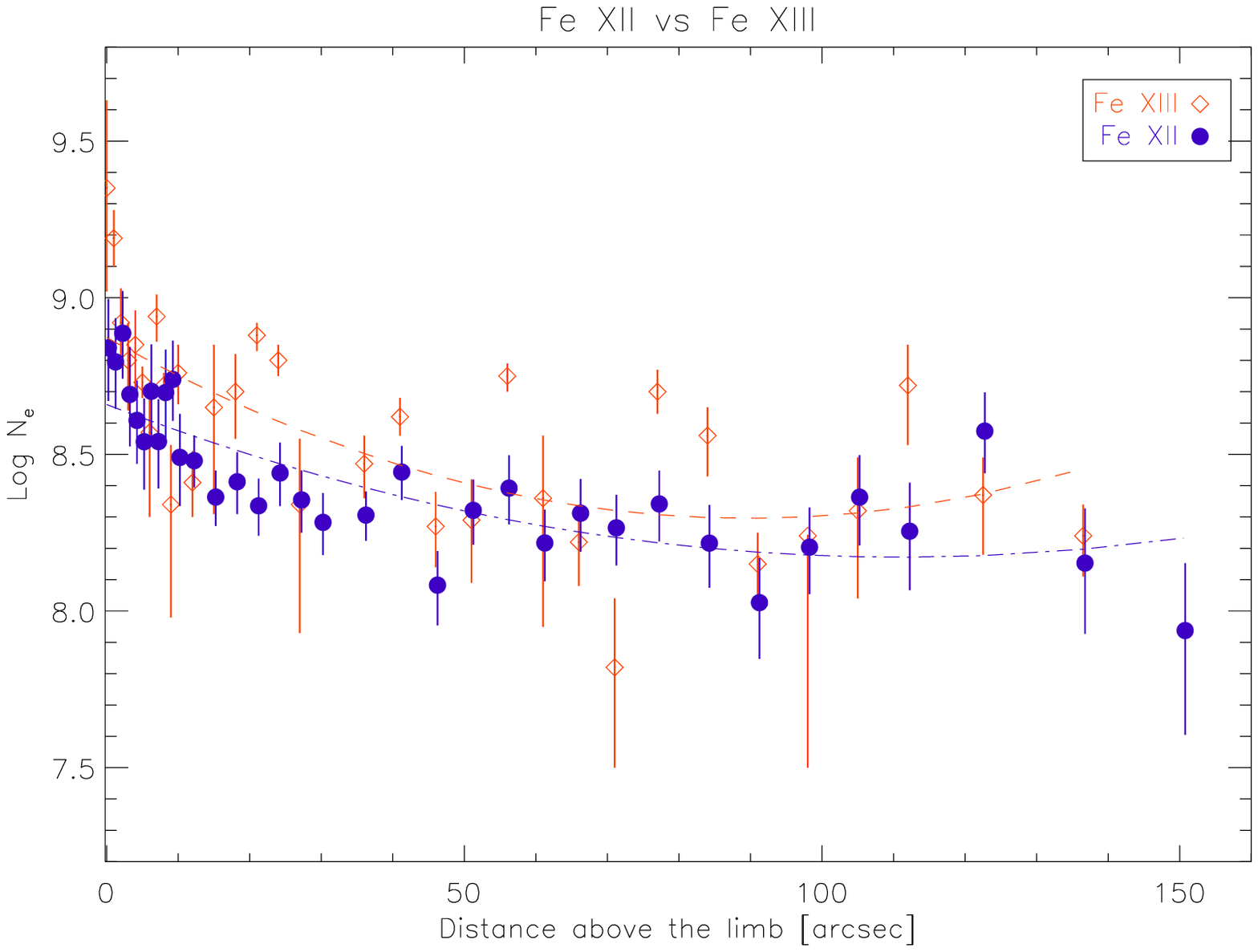}
\includegraphics[bb= 0 0 520 340, clip= true, width=8cm]{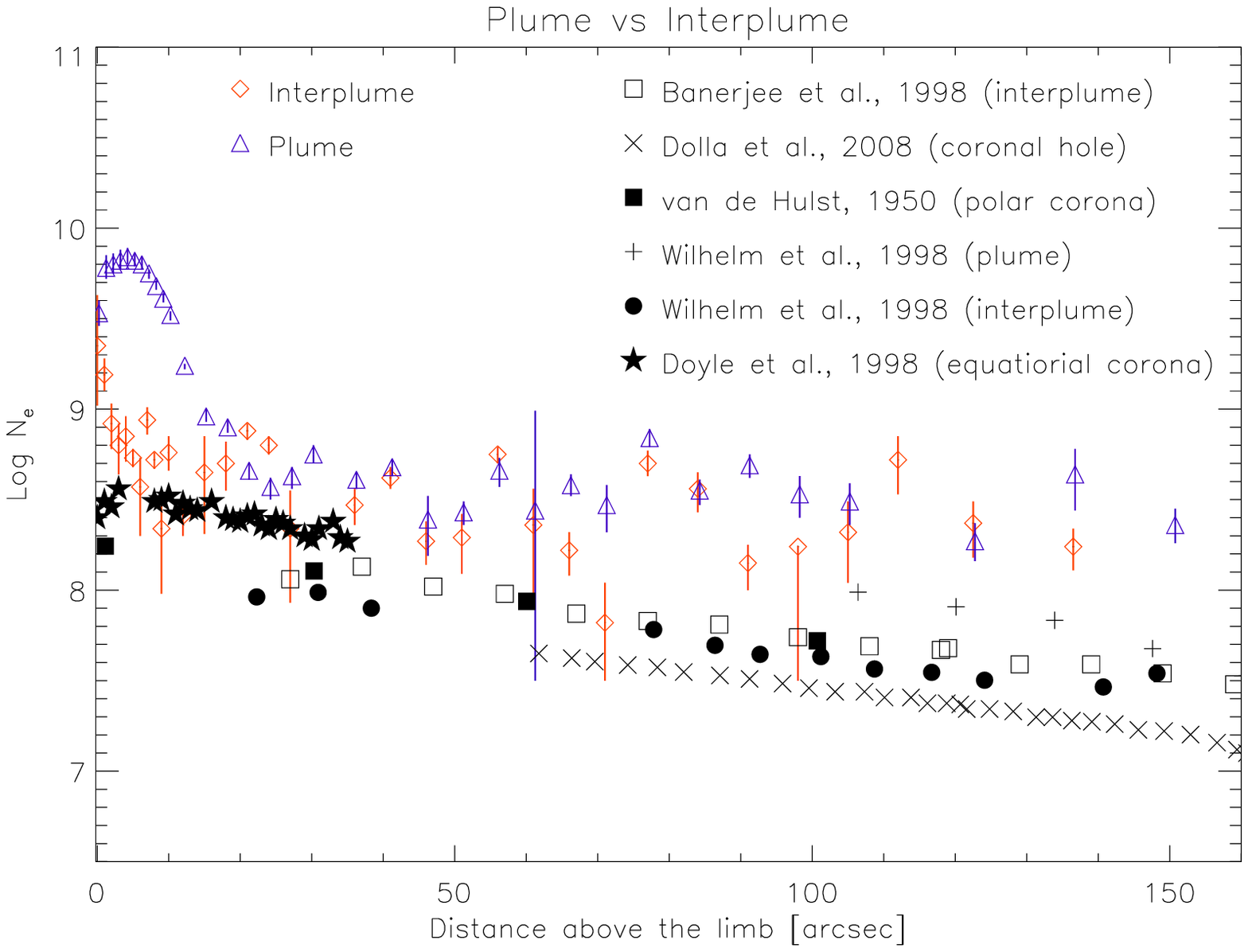}
\caption {Variation in density with height in the polar coronal hole region. The top panel shows results  as recorded by \ion{Fe}{xiii}, along plume and interplume regions. The middle panel shows the comparison of \ion{Fe}{xii} and \ion{Fe}{xiii}. The lower panel shows a comparison of density values as recoded by different instruments within the same radial distance.} \label{fig:densitycomp}
\end{figure}
%%%%%%%%%%%%%%%%%%%%%%%%%%%%%%%%%%%%%%%%%%%%%%%%%%%%%%%%%%%%%%%%
%%%%%%%%%%%%%%%%%%%%%%%% FIG comparison with SUMER %%%%%%%%%%%%%%%%%%%%%%%
\begin{figure}
\centering
\includegraphics[width=8cm]{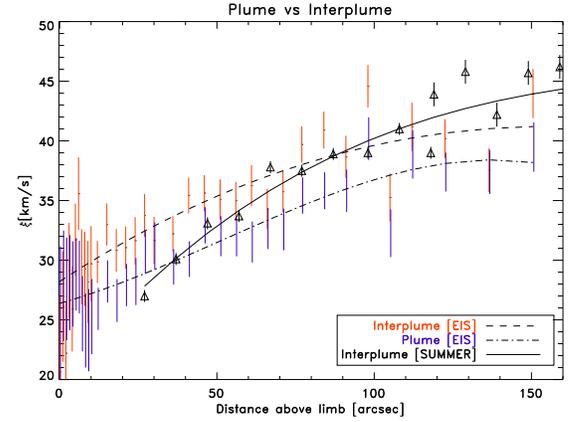}
\caption {Variation in nonthermal velocity with height as recorded by 
Fe~{\sc xii} 195~\AA\ along a polar plume and interplume (the present study). 
The solid line corresponds to the nonthermal velocity as derived from  
Si~{\sc viii} 1445.75\AA\ \citep{1998A&A...339..208B}. The dashed 
line is a third-order polynomial fit. } \label{fig:comparesumer}
\end{figure}
%%%%%%%%%%%%%%%%%%%%%%%%%%%%%%%%%%%%%%%%%%%%%%%%%%%%%%%%%%%%%%%%
Fig.~\ref{fig:map} shows maps of line intensity and FWHM  from \ion{Fe}{xii}~195.12~\AA\  
and electron density (from \ion{Fe}{xii}) of the northern polar off-limb coronal hole region. 

Observations have  revealed that plasma conditions in polar plumes are quite different from interplume lanes 
\citep{1997SoPh..175..375H, 1998ApJ...500.1023W}. From these maps it 
is clear that  plumes have slightly higher density and lower FWHM. Closer to the 
limb at the base of the plume (around $X=15$) one also finds the presence of a coronal bright point and it does affect the density values close to the limb. To look for a possible
correlation/anti-correlation between the line intensity and the most probable 
speed, we concentrate on specific heights. As we go off the limb, 
Fig.~\ref{fig:Ycut} shows the line intensity and the FWHM along three strips tangent to the limb at different heights. We find 
 evidence of an anti-correlation between the intensity and FWHM. If we go 
out farther away from the limb, the anti-correlation is weaker, because at 
these heights above the limb the plume structures have expanded slightly 
nonradially (see Fig.~\ref{fig:map}). This observed anti-correlation between 
intensity and width in polar plumes have been reported by 
\cite{1997ESASP.404..175A} and \cite{1997AdSpR..20.2219N} with the UVCS 
instrument and also by  \cite{1997SoPh..175..375H} and 
\cite{2000SoPh..194...43B} with SUMER. Now, to study the variations of different 
parameters namely density and FWHM with height, we focus our attention on a 
fixed location, $X=15$, as a representative location for the plume and $X = -57$, as a 
representative case for the interplume. 

In Fig.~\ref{fig:densitycomp}  we plot the variation of the density  with height as recorded by \ion{Fe}{xiii} (top panel) along  plume and interplume (along the two dashed vertical lines as drawn in Fig.~\ref{fig:Ycut})\footnote{ 
A summary of the measured parameters and calculated values are tabulated in the 
on-line material,  Table.~\ref{table} for \ion{Fe}{xii} and \ion{Fe}{xiii}.}. The dashed lines in Fig.~\ref{fig:densitycomp}
correspond to a second order polynomial fit. One can clearly see that at the 
solar limb, the plume density is almost one order of magnitude higher than the 
interplume, but around 70\arcsec\  off-limb, the densities of the plume and 
interplume seem to be very close. The errors in deriving the 
electron density in Fig.~\ref{fig:densitycomp} depends not only on the line 
strengths but also on the uncertainty within the atomic data, which can be upto 20\% \citep{2005A&A...433..731D}. We should point out here that, for polar regions, these densities are higher than the previously published  values.  To validate our measurements, we also calculated the densities using \ion{Fe}{xii} lines. For the same range of heights and along the same interplume lane, we compare the results in the middle panel of Fig.~\ref{fig:densitycomp}. It clearly shows that the densities are consistent (within their error bars) as recorded by two different line pairs of \ion{Fe}{xiii} and \ion{Fe}{xii}. In the lower panel of Fig.~\ref{fig:densitycomp} we show a detailed comparison of our measured densities with earlier published values for the same range of heights  for different coronal conditions \citep{1998ApJ...500.1023W,1950BAN....11..135V,2008A&A...483..271D}. 

Our densities are marginally comparable to previously published values for plumes and for the quiet corona. This may stem form the new atomic data for \ion{Fe}{xiii} and \ion{Fe}{xiii} used in this analysis and/or contamination from the background quite sun.
Assuming ionisation equilibrium and using Eq. (1), we now calculate the 
nonthermal speed and study its variation with height (see Fig.~\ref{fig:comparesumer}) for the \ion{Fe}{xii} (as this is the strongest line and line widths are measured with better accuracies). 
The variation in the nonthermal velocity  with height along 
the plume and interplume reveals that, consistently with height, the nonthermal 
velocity  is slightly higher in the interplume than the plume.
%For the plume, the nonthermal 
%velocity 
%increases from $\sim 40 ~{\rm km ~s}^{-1}$ at 25\arcsec\ above the limb to 
%$\sim 50~{\rm km ~ s}^{-1}$ at 150\arcsec\ above the limb for plasma around 
%$ 1.3\times10^6$ K. For the interplume, the nonthermal velocity increase from 
%$\sim 42~{\rm km ~s}^{-1}$ at 17\arcsec\ above the limb to 
%$\sim 52~{\rm km ~ s}^{-1}$ at 150\arcsec\ above the limb. 
For comparison with previously reported  results, we overplot the nonthermal velocity as derived 
from Si~{\sc viii}~1445.75\AA\ \citep{1998A&A...339..208B} as 
represented by triangles along with their best fit in Fig.~\ref{fig:comparesumer}.
%%%%%%%%%%%%%%%%%%%%%%%% FIG density and FWHM relation for interplume and plume %%%%%%%%%%%%%%%%%%%%%%%%%%%%%%%%%%%%%%%%%
\begin{figure}
\centering
\includegraphics[width=8cm]{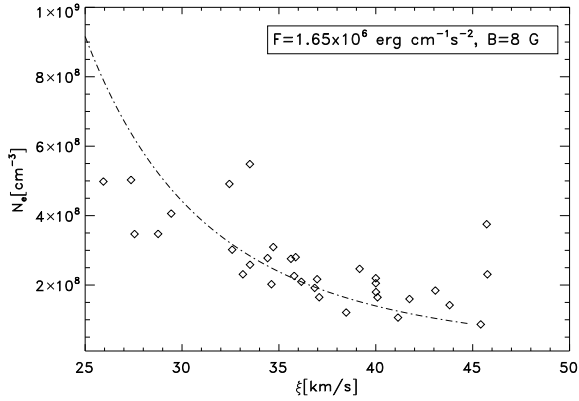}
\caption {Variation in electron density with nonthermal velocity for the
polar coronal hole. The squared boxes represents 
the observed values and the solid line represents the theoretical relation 
(Eq.~[4]) for fixed magnetic field strength as indicated.} \label{fig:relation}
\end{figure}
%%%%%%%%%%%%%%%%%%%%%%%%%%%%%%%%%%%%%%%%%%%%%%%%%%%%%%%%%%%%%%%%

\section{Discussion and conclusions} 
Using the EIS spectrometer on Hinode, we have studied the variations in the line width and the 
electron density as a function of height. We used the density sensitive line pairs of Fe~{\sc xii}  and Fe~{\sc xiii}, and   calculated  densities have comparable values (within their error bars). We must point out that the densities as reported here are consistently higher than previously published values (see Fig.~\ref{fig:densitycomp}). The difference may stem from real physical
 differences between the emitting regions of the coronal hole plasma around November 2007 and previous coronal conditions as recorded by other instruments (e.g. SUMER). Other factors could be the CHIANTI atomic models. There has been a recent revision in the \ion{Fe}{xiii} atomic model \citep{2009ApJ...692.1294W}, and we used the updated CHIANTI database for our calculation. We should also point out that, from the full scan image of the density (Fig.~\ref{fig:densitymapfull}, available in the online version only), it shows that the off-limb coronal densities are close to the quiet Sun values and in fact somewhat higher than the on-disk coronal hole densities. This suggests that the coronal emission may have been affected by background emission from the quiet coronal structures.

Alfv\'en waves propagate virtually undamped through the quasi-static corona and 
deposit their energy flux in the higher corona. The total energy flux crossing 
a surface of area $A$ in the corona due to Alfv\'en waves is given by 
\citep{2001A&A...374L...9M}
\begin{equation}
F = \sqrt{\frac{\rho}{4 \pi}} <\delta v^2> B\ A ~~, 
\end{equation}
where $\rho$ is the plasma mass density (related to $N_e$ as $\rho = m_p N_e$,
$m_p$ is the proton-mass), $<\delta v^2>$ is the mean square velocity related 
to the observed nonthermal velocity as $ \xi^2 = <\delta v^2>/2 $, and 
$B$, is the magnetic field strength. If the wave energy flux is conserved as 
the wave propagate outwards, then Eq.(2) gives
\begin{equation}
<\delta v^2>^{1/2} \ \propto \ \rho^{-1/4} (BA)^{-1/2}~~,
\end{equation}
Now if one assumes a flux tube geometry, then $BA$ is constant with height and 
Eq.(3) yields 
\begin{equation}
<\delta v^2>^{1/2} \ \propto \ \rho^{-1/4}~~.
\end{equation}
From our dataset (for \ion{Fe}{xii} 195 \AA\ ) at 30\arcsec\ above the limb for the interplume location (see Table.~\ref{table}) 
using $N_e = 2.2 \times 10^8 ~{\rm cm}^{-3}$, $<\delta v^2>  = 2 \times 
(43.7 ~{\rm km~s}^{-1})^{2}$, we find $ F = 1.85 \times 10^6$ ergs cm$^{-2}$ 
s$^{-1}$ for $B = 8$ G, which is only slightly higher than the requirements 
for a coronal hole with a high-speed solar wind flow 
\citep{1977ARA&A..15..363W}, which is in turn estimated to be $8 \times 10^5$ ergs 
cm$^{-2}$ s$^{-1}$. The average field strength in coronal holes is estimated to 
be 5-10 G \citep{1990CoPhR..12..205H, 2008ApJ...688.1374T}.

To explore the validity of Eq.(4),  we plot Fig.~\ref{fig:relation} corresponding to our interplume data. The solid lines are the theoretically predicted functional form of 
the variation of number density with nonthermal velocity (Eq.[4]) and the 
diamonds are our observed data. The proportionality constant have been chosen 
to match the calculated energy flux. For the interplume data we used B=8 G. 
Once again the agreement is very good, especially when we are away from 
the limb. 
%Note that for the plume data-points in Fig.~\ref{fig:relation} we 
%seem to have two groups of data. To fit the observed data corresponding to 
%close to limb locations (the higher density values) for the same value 
%of magnetic field, a slightly higher energy flux is required. We have checked these points and they 
%correspond to locations where we see loop like structures very close to the limb 
%with strong velocity flows. We conjecture that these smaller magnetic 
%structures may be isolated from the larger expanding tube. 
 
Our observations provide observational signatures for the presence of Alfv\'en 
waves in polar coronal regions at least within 1.1$R_{\odot}$. These upwardly 
propagating Alfv\'en waves may heat the corona and accelerate the solar wind. 
The slightly larger widths within the interplume regions as compared to plumes also 
indicate that probably interplumes are the preferred channel for the 
acceleration of the wind. In this dataset we do not find large differences between plumes and interplumes beyond 70\arcsec above the limb. Finally the EIS line width results are consistent with previous results from SUMER.

\onltab{1}{
\begin{table*}
\caption{A summary of measured intensity, calculated nonthermal
 velocity ($\xi$), flux ratio and electron density for the plume (P) and interplume
 (I-P) obtained from \ion{Fe}{xii} ($T_{i}=1.3\times10^{6}$K) and
 \ion{Fe}{xiii} ($T_{i}=1.6\times10^{6}$K) at different locations.}
\begin{tabular}{c c c c c c c c c c c c c c c c c}\hline\hline
Above limb & \multicolumn{2}{c}{I(195.12\AA)} & \multicolumn{2}{c}{$\xi$} & \multicolumn{2}{c}{\ion{Fe}{xii} ratio} & \multicolumn{2}{c}{$\log N_{e}/cm^{-3}$} & \multicolumn{2}{c}{I(202.04\AA)} & \multicolumn{2}{c}{$\xi$} & \multicolumn{2}{c}{\ion{Fe}{xiii} ratio} & \multicolumn{2}{c}{$\log N_{e}/cm^{-3}$} \\
(arcsec) & \multicolumn{2}{c}{erg cm$^{2}$s$^{-1}$sr$^{-1}$} & \multicolumn{2}{c}{km$\,$s$^{-1}$} & \multicolumn{2}{c}{$10^{-2}$}& & & \multicolumn{2}{c}{erg cm$^{2}$s$^{-1}$sr$^{-1}$} & \multicolumn{2}{c}{km$\,$s$^{-1}$} &\multicolumn{2}{c}{$10^{-2}$}& & \\
& I-P & P & I-P & P & I-P & P & I-P & P & I-P & P & I-P & P & I-P & P & I-P & P \\ \hline
   0.00 &  52.45 & 164.67 & 34.6 & 36.9 & 20.25 & 22.71 & 8.84 & 8.95 &  34.35 &  93.85 & 20.7 & 24.9 & 50.52 & 64.70 & 9.35 & 9.53 \\ 
   1.00 &  62.02 & 235.34 & 29.3 & 38.4 & 19.27 & 24.54 & 8.80 & 9.03 &  41.81 & 137.62 & 26.1 & 27.9 & 38.72 & 84.76 & 9.19 & 9.78 \\ 
   2.00 &  69.19 & 338.49 & 28.3 & 39.6 & 21.29 & 25.57 & 8.89 & 9.07 &  48.19 & 190.09 & 22.2 & 27.6 & 22.87 & 86.29 & 8.92 & 9.80 \\ 
   3.00 &  75.52 & 466.16 & 29.7 & 35.1 & 17.08 & 26.58 & 8.69 & 9.11 &  52.97 & 249.19 & 29.2 & 28.1 & 17.85 & 87.73 & 8.80 & 9.82 \\ 
   4.00 &  78.00 & 591.03 & 26.7 & 33.5 & 15.42 & 27.77 & 8.61 & 9.15 &  57.13 & 304.34 & 25.9 & 28.0 & 19.98 & 88.91 & 8.85 & 9.84 \\ 
   5.00 &  77.34 & 675.66 & 24.7 & 29.2 & 14.12 & 29.85 & 8.54 & 9.23 &  60.63 & 341.56 & 31.7 & 28.8 & 15.58 & 87.72 & 8.73 & 9.82 \\ 
   6.00 &  77.36 & 723.83 & 24.5 & 27.3 & 17.29 & 31.22 & 8.70 & 9.28 &  64.50 & 361.62 & 35.6 & 28.6 & 10.73 & 86.09 & 8.57 & 9.80 \\ 
   7.00 &  79.45 & 706.95 & 26.0 & 25.5 & 14.13 & 32.34 & 8.54 & 9.32 &  65.34 & 363.00 & 29.4 & 25.5 & 23.97 & 82.00 & 8.94 & 9.75 \\ 
   8.00 &  79.13 & 664.77 & 23.0 & 25.2 & 17.20 & 32.43 & 8.70 & 9.32 &  65.01 & 347.64 & 29.3 & 24.1 & 15.04 & 77.39 & 8.72 & 9.68 \\ 
   9.00 &  81.20 & 592.35 & 30.8 & 25.1 & 18.07 & 31.95 & 8.74 & 9.31 &  66.42 & 318.12 & 28.2 & 24.1 &  6.45 & 71.80 & 8.34 & 9.61 \\ 
  10.00 &  80.88 & 505.33 & 32.0 & 24.1 & 13.21 & 31.79 & 8.49 & 9.30 &  67.35 & 277.62 & 29.7 & 25.3 & 16.50 & 63.80 & 8.76 & 9.52 \\ 
  12.00 &  80.63 & 339.73 & 29.9 & 22.1 & 13.02 & 29.52 & 8.48 & 9.22 &  66.44 & 183.97 & 29.8 & 25.9 &  7.67 & 42.09 & 8.41 & 9.24 \\ 
  15.00 &  78.40 & 201.39 & 30.5 & 28.4 & 11.05 & 19.71 & 8.36 & 8.82 &  65.57 & 125.56 & 33.0 & 28.2 & 12.94 & 25.23 & 8.65 & 8.96 \\ 
  18.00 &  73.91 & 165.77 & 30.8 & 30.4 & 11.85 & 15.40 & 8.41 & 8.61 &  67.02 & 113.20 & 30.8 & 26.6 & 14.58 & 22.32 & 8.70 & 8.90 \\ 
  21.00 &  71.94 & 144.86 & 34.3 & 29.4 & 10.61 & 12.68 & 8.34 & 8.46 &  61.95 & 103.00 & 31.1 & 27.9 & 21.06 & 13.27 & 8.88 & 8.66 \\ 
  24.00 &  68.51 & 131.47 & 33.0 & 27.3 & 12.33 & 13.57 & 8.44 & 8.51 &  60.72 &  97.82 & 31.6 & 28.2 & 18.04 & 10.72 & 8.80 & 8.57 \\ 
  27.00 &  66.44 & 121.21 & 33.1 & 28.4 & 10.90 & 11.48 & 8.35 & 8.39 &  59.08 &  95.80 & 33.7 & 30.7 &  6.55 & 12.26 & 8.34 & 8.63 \\ 
  30.00 &  63.69 & 116.30 & 34.2 & 27.0 &  9.80 & 10.66 & 8.28 & 8.34 &  56.08 &  89.60 & 31.6 & 31.2 &  0.70 & 16.14 & 7.50 & 8.75 \\ 
  36.00 &  57.09 & 100.84 & 31.9 & 29.7 & 10.15 & 11.06 & 8.31 & 8.36 &  52.96 &  81.91 & 32.2 & 29.4 &  8.66 & 11.76 & 8.47 & 8.61 \\ 
  41.00 &  53.43 &  93.63 & 31.7 & 27.3 & 12.38 &  9.21 & 8.44 & 8.24 &  50.69 &  76.62 & 35.4 & 30.1 & 12.21 & 13.79 & 8.62 & 8.68 \\ 
  46.00 &  52.43 &  86.73 & 35.8 & 28.1 &  7.16 &  9.12 & 8.08 & 8.24 &  48.36 &  71.34 & 35.6 & 32.9 &  5.60 &  7.19 & 8.27 & 8.39 \\ 
  51.00 &  48.57 &  81.15 & 33.5 & 33.0 & 10.38 & 10.60 & 8.32 & 8.34 &  46.97 &  70.03 & 35.1 & 32.0 &  5.76 &  8.00 & 8.29 & 8.43 \\ 
  56.00 &  46.51 &  74.57 & 36.5 & 32.6 & 11.52 &  9.66 & 8.39 & 8.27 &  43.59 &  61.55 & 35.0 & 31.9 & 16.10 & 13.13 & 8.75 & 8.66 \\ 
  61.00 &  42.88 &  65.98 & 34.4 & 31.4 &  8.86 & 11.49 & 8.22 & 8.39 &  40.57 &  57.00 & 36.3 & 31.5 &  6.72 &  8.19 & 8.36 & 8.44 \\ 
  66.00 &  39.34 &  61.77 & 37.4 & 32.6 & 10.24 & 10.56 & 8.31 & 8.33 &  39.13 &  52.93 & 34.3 & 32.6 &  4.94 & 11.07 & 8.22 & 8.58 \\ 
  71.00 &  40.04 &  57.45 & 40.4 & 33.3 &  9.55 &  8.65 & 8.27 & 8.20 &  36.33 &  50.11 & 35.7 & 32.6 &  2.08 &  8.64 & 7.82 & 8.47 \\ 
  77.00 &  36.04 &  52.79 & 37.4 & 34.8 & 10.70 &  9.90 & 8.34 & 8.29 &  34.49 &  46.50 & 39.7 & 35.4 & 14.54 & 19.55 & 8.70 & 8.84 \\ 
  84.00 &  31.71 &  46.21 & 37.4 & 32.6 &  8.86 &  9.20 & 8.22 & 8.24 &  29.95 &  41.55 & 40.9 & 35.8 & 10.53 & 10.27 & 8.56 & 8.55 \\ 
  91.00 &  29.47 &  41.63 & 38.5 & 35.8 &  6.55 &  6.95 & 8.03 & 8.06 &  26.83 &  36.19 & 38.6 & 35.8 &  4.22 & 14.17 & 8.15 & 8.69 \\ 
  98.00 &  26.06 &  36.87 & 39.1 & 35.6 &  8.68 &  9.05 & 8.20 & 8.23 &  24.20 &  32.70 & 44.6 & 40.2 &  5.23 &  9.89 & 8.24 & 8.53 \\ 
 105.00 &  24.47 &  34.22 & 43.1 & 34.1 & 11.04 & 10.78 & 8.36 & 8.35 &  22.11 &  29.07 & 35.3 & 32.3 &  6.19 &  9.02 & 8.32 & 8.49 \\ 
 112.00 &  23.00 &  31.14 & 37.4 & 34.6 &  9.39 &  6.49 & 8.25 & 8.02 &  21.09 &  29.55 & 41.2 & 38.9 & 14.96 & -2.76 & 8.72 & **** \\ 
 122.50 &  20.49 &  28.04 & 43.0 & 39.0 & 14.76 &  9.86 & 8.57 & 8.29 &  17.72 &  24.00 & 40.2 & 37.4 &  6.88 &  5.61 & 8.37 & 8.27 \\ 
 136.50 &  16.80 &  22.50 & 41.1 & 39.8 &  8.02 &  5.26 & 8.15 & 7.89 &  14.61 &  19.07 & 37.5 & 37.4 &  5.24 & 12.69 & 8.24 & 8.64 \\ 
 150.50 &  14.68 &  18.72 & 42.7 & 40.3 &  5.66 &  8.82 & 7.94 & 8.21 &  12.31 &  15.80 & 44.0 & 39.5 &  0.05 &  6.85 & 7.50 & 8.36 \\ 
\hline\label{table}
\end{tabular}
\end{table*}
}
%%%%%%%%%%%%%%%%%%%%%%%% FIG density map full raster scan%%%%%%%%%%%%%%%%%%%%%%%%%%%%%%%%%%%%%%%%%
\onlfig{7}{
\begin{figure*}
\centering
\includegraphics[bb= 5 45 520 360, clip= true, width=10cm]{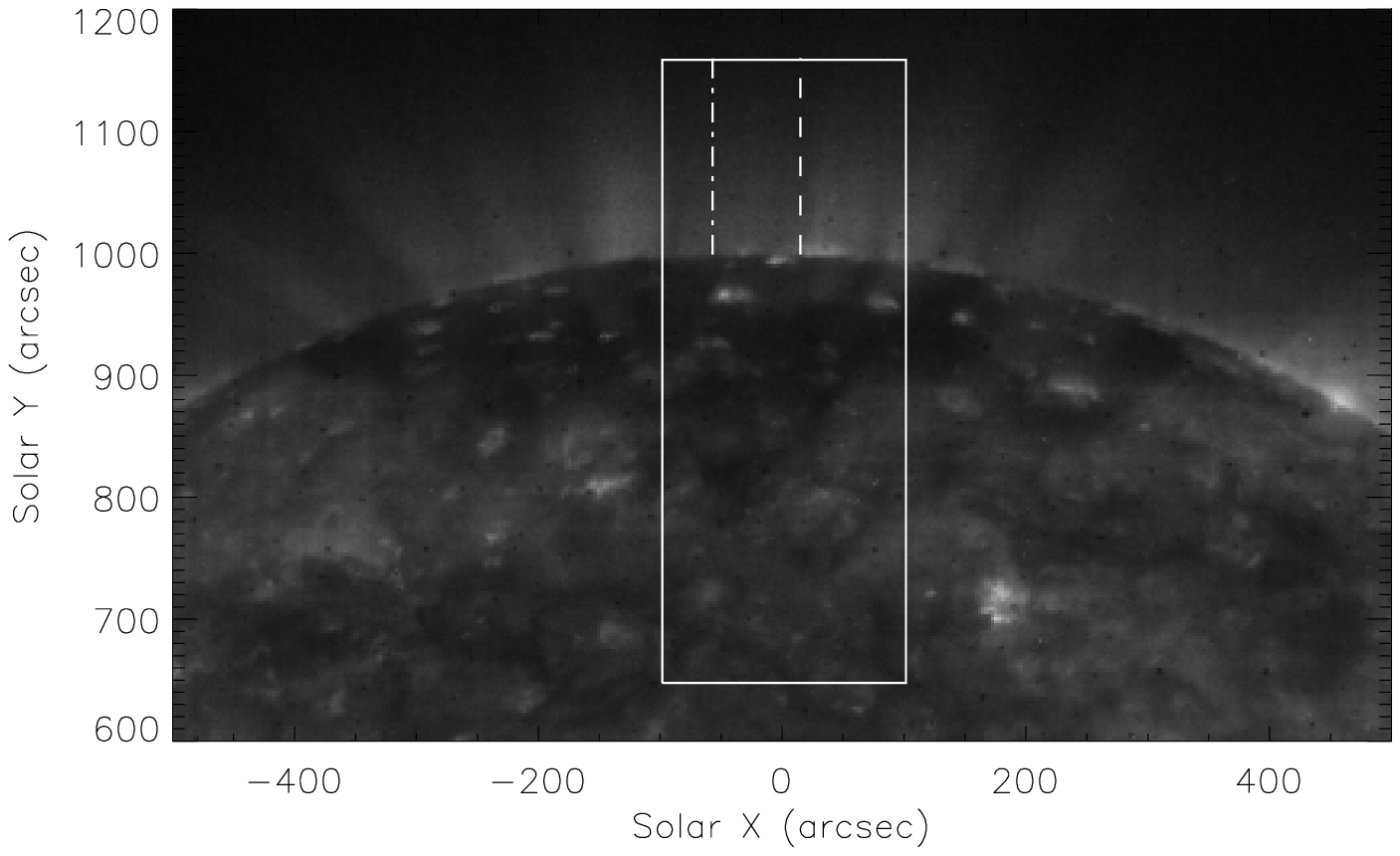}
\hspace*{-3cm}\includegraphics[bb= 5 5 520 360, clip= true, width=8cm]{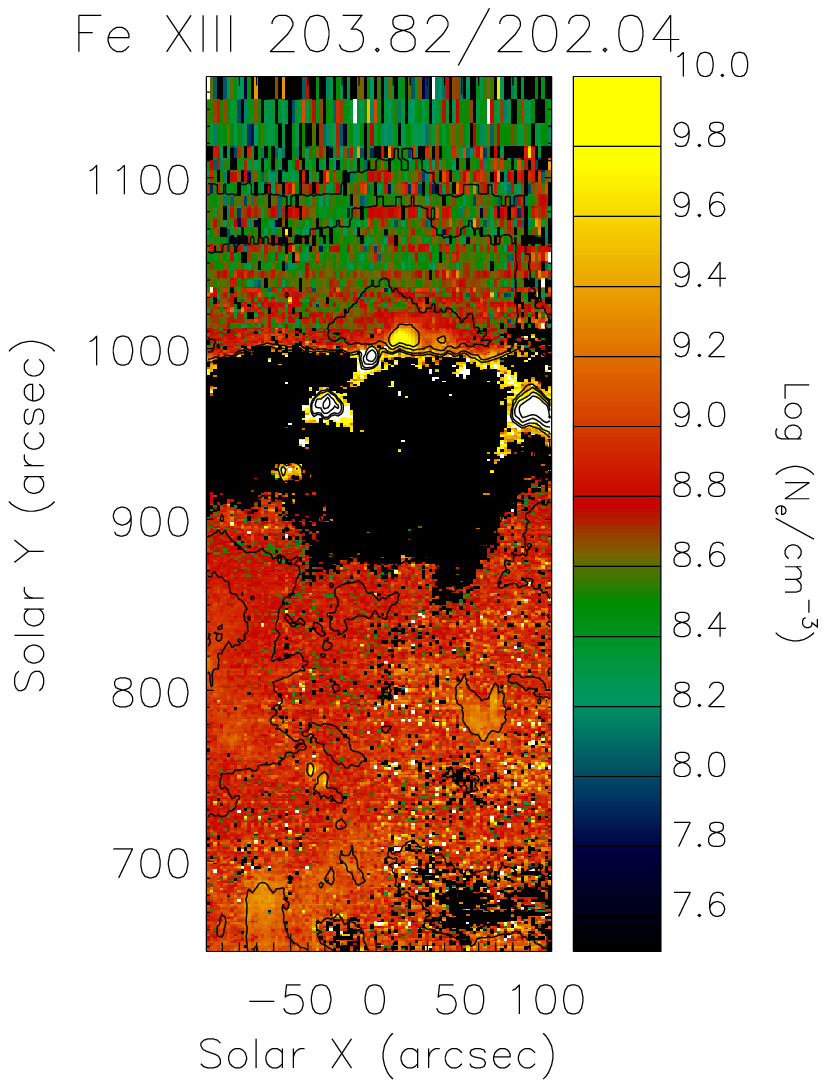}
\caption {The left panel shows the EIT 171~\AA\ image with EIS field of view overplotted (square
 box) and plume and interplume chosen. The right panel shows the density map corresponding to the full EIS raster scan as recorded by \ion{Fe}{xiii} and the contours show the intensity. \label{fig:densitymapfull}}
\end{figure*}
}
%%%%%%%%%%%%%%%%%%%%%%%% FIG Comp Non-thermal Vel Fe XII and XIII %%%%%%%%%%%%%%%%%%%%%%%%%%%%%%%%%%%%%%%%%
%\onlfig{7}{
%\begin{figure}
%\centering
%\includegraphics[bb= 5 45 520 360, clip= true, width=8cm]{071010OL_VelCompYx21x57.eps}
%\caption {Comparison of nonthermal velocities for \ion{Fe}{xii} and
% \ion{Fe}{xiii} for the plume area.} \label{fig:Nonthvelcomp}
%\end{figure}
%}
%%%%%%%%%%%%%%%%%%%%%%%%%%%%%%%%%%%%%%%%%%%%%%%%%%%%%%%%%%%%%%%%%%%%%%%%%%%%%%
\begin{acknowledgements}                 
Research at Armagh Observatory is grant-aided
by the N.~Ireland Dept. of Culture, Arts, and Leisure.  This work 
was supported by a Royal Society/British Council and  STFC grant PP/D001129/1.
The SUMER project is financially supported by DLR, CNES, NASA, and the ESA 
PRODEX programme (Swiss contribution). We thank the referee for valuable suggestions, which has improved the quality of this article. 
\end{acknowledgements}
%%%%%%%%%%%%%%%% References %%%%%%%%%%%%%%%%%%%%%%%%%%%%%%
\bibliographystyle{aa.bst}
\bibliography{banerjee}
\end{document}